# The GDPR's Rules on Data Breaches: Analysing Their Rationales and Effects

*Frederik Zuiderveen Borgesius*, Hadi Asghari**,*

*Noël Bangma***, Jaap-Henk Hoepman*****

**Abstract**

The General Data Protection Regulation (GDPR) requires an organisation that suffers a data breach to notify the competent Data Protection Authority. The organisation must also inform the relevant individuals, when a data breach threatens their rights and freedoms. This paper focuses on the following question: given the goals of the GDPR's data breach notification obligation, what are its strengths and weaknesses? We identify six goals of, or rationales for, the GDPR`s data breach notification obligation, and we assess the obligation in the light of those goals. We refer to insights from information security and economics, and present them in a reader-friendly way for lawyers. Our main conclusion is that the GDPR's data breach rules are likely to contribute to the goals. For instance, the data breach notification obligation can nudge organisations towards better security; such an obligation enables regulators to perform their duties; and such an obligation improves transparency and accountability. However, the paper also warns that we should not have unrealistic expectations of the possibilities for people to protect their interests after a data breach notice. Likewise, we should not have high expectations of people switching to other service providers after receiving a data breach notification. Lastly, the paper calls for Data Protection Authorities to publish more information about reported data breaches. Such information can help to analyse security threats.

---

* Prof. dr. Frederik Zuiderveen Borgesius, prof. ICT and law, iHub, the interdisciplinary research hub on digitalization and society, Radboud University, Nijmegen, The Netherlands, frederikzb@cs.ru.nl. Frederik has received funding from the European Union's Horizon Europe research and innovation program under grant agreement No 101070212, for the FINDHR project.
** Dr. Hadi Asghari, senior researcher at the Alexander von Humboldt Institute for Internet and Society, Berlin, Germany, hadi.asghari@hiig.de.
*** Noël Bangma, Research Intern at iHub, noel@familiebangma.nl.
**** Prof. dr. Jaap-Henk Hoepman, Radboud University, University of Groningen and Karlstad University, jhh@cs.ru.nl.

# 1 Introduction

'Each year brings more bad news. Each year steals the crown for the most awful year for data breaches,' write Solove and Hartzog, discussing mostly the situation in the US. They speak of a 'data breach epidemic'.[1] Perhaps they should have said a data breach 'pandemic', as data breaches are a problem all over the world. The General Data Protection Regulation (GDPR) requires an organisation that suffers a data breach to notify the competent Data Protection Authority (DPA). The organisation must also inform the relevant individuals, when a data breach threatens their interests.

The paper focuses on the following question: given the goals of the GDPR's data breach notification obligation, what are its strengths and weaknesses? The paper contributes to scholarship is several ways. First, building on earlier literature, we propose a taxonomy of six goals of, or rationales for, the GDPR`s data breach notification obligation. Second, we assess the obligation in the light of those goals. Third, we combine insights from different disciplines (law, information security, and economics). We aim to present insights from information security and economics in a reader-friendly way for lawyers. Fourth, we aim to describe the GDPR's data breach rules in a reader-friendly way for readers who are not specialised in the GDPR.

The paper could be relevant for policymakers, within and outside the EU, who consider adopting or amending notification obligations regarding data breaches. The paper could also be relevant for data subjects and practitioners, such as lawyers and data protection officers who are working with the GDPR's provisions. Non-compliance with the rules can lead to large fines.[2]

---

1 Daniel Solove and Woodrow Hartzog, *Breached!: Why Data Security Law Fails and How to Improve it* (OUP 2022), 44.

2 Article 83 of the GDPR.

The paper is structured as follows: In section 2, we explain the GDPR rules on notifying data breaches and their background. Readers who know those rules by heart could skip this section. In section 3, we identify six rationales for, or goals of, the data breach notification obligation. We also analyse whether the obligation is likely to reach these goals. Lastly, we provide suggestions for further research and a conclusion (section 4 and 5).

# 2 The GDPR and data breaches

## 2.1 Background

Since the early 1970s, data protection laws in Europe emphasise the importance of security and confidentiality of personal data.[3] The predecessor of the GDPR, the Data Protection Directive from 1995, contained security obligations,[4] but did not contain a data breach notification obligation.

The first law to impose a reporting obligation for data leaks was the California Security Breach Information Act of 2003. Now, all states in the United

3 See for instance the Data Protection Act of the German state of Hessen (G V Bl.II 300-10, published at Wiesbaden, 12 October 1970, in Gesetz-und Verordnungsblatt für das Land Hessen [Laws and Regulations Journal], Part I, No . 41), Article 2: 'The records, data and results covered by data protection shall be obtained, transmitted and stored in such a way that they cannot be consulted, altered, extracted or destroyed by an unauthorized person. This shall be ensured by appropriate staff and technical arrangements.' See also Article 3. Security was also mentioned in Committee of Ministers, Resolution (73)22 on the protection of the privacy of individuals vis-à-vis electronic data banks in the private sector, 26 September 1973, par 8 and 9, and in Committee of Ministers, Resolution (74)29 on the protection of the privacy of individuals vis-à-vis electronic data banks in the public sector, 20 September 1974, par. 6 and 7. See for a history of the security requirements in European data protection law: J.A. Hofman, 'De beveiliging van persoonsgegevens. Over de juridische invulling van art. 5 lid 1 onder f en 32 AVG' ['The security of personal data. About the legal interpretation of article 5(1)(f) and article 32 GDPR'] (DPhil thesis, Radboud University 2022), Chapter 4.

4 Article 17 of Directive 95/46/EC of the European Parliament and of the Council of 24 October 1995 on the protection of individuals with regard to the processing of personal data and on the free movement of such data ('Data Protection Directive').

States have a data breach notification obligation.[5] In the EU, the ePrivacy Directive introduced a data breach notification in 2009, which applied, roughly summarised, to telecommunications providers.[6]

The GDPR, adopted in 2016 and applicable in the EU since 2018, includes an 'integrity and confidentiality' principle among its seven overarching principles. The GDPR's integrity and confidentiality principle obliges controllers to ensure 'appropriate security' for personal data, 'including protection against unauthorised or unlawful processing and against accidental loss, destruction or damage, using appropriate technical or organisational measures.'[7] A controller is, roughly summarised, an organisation that uses personal data.[8] The GDPR also includes data breach notification obligations; we turn to those now.

## 2.2 The GDPR and data breaches

The GDPR's obligation to notify data breaches can be summarised as follows. There are two data breach notification obligations in the GDPR: one to the Data Protection Authority (DPA) (article 33) and one to the data subject (article 34). Under article 33, a controller must report a data breach to the national DPA, unless the breach is unlikely to result in risks for people. Under article 34, a controller must notify a data breach to the data subject, when the breach is likely to result in a high risk to the rights and freedoms of the data subject. (The data

---

[5] Solove and Hertzog (n 1).

[6] Article 4 of the Directive 2002/58/EC of the European Parliament and of the Council of 12 July 2002 concerning the processing of personal data and the protection of privacy in the electronic communications sector (as amended by Directive 2009/136/EC of the European Parliament and of the Council of 25 November 2009) ('ePrivacy Directive'). See in detail about that provision: Rosa Barcelo and Peter Traung, 'The emerging European union security breach legal framework: The 2002/58 ePrivacy Directive and beyond' in Serge Gutwirth et. al (eds), *Data Protection in a Profiled World* (Springer 2010).

[7] Article 5(1)(f) of the GDPR; See also recital 39 of the GDPR.

[8] A controller is defined in Article 4(7) of the GDPR.

subject is the person whose personal data are processed.[9])

Below, we summarise the main features of the GDPR's data breach notification obligation. We refer to the GDPR's provisions, and to the GDPR's 'preamble'. European Union statutes start with a preamble, which consists of a list of 'recitals'. Recitals are not legally binding on their own; they can only be used to interpret the GDPR's provisions.[10] We also refer to guidance by the European Data Protection Board (EDPB).[11] The EDPB is an independent European body, that aims to contribute to a consistent application of data protection rules throughout the EU. The DPAs of the EU Member States cooperate in the EDPB. The EDPB is the successor of the Article 29 Working Party.[12]

## 2.3 What counts as a data breach?

What counts as a data breach in the GDPR? The GDPR defines a 'personal data breach' as 'a breach of security leading to the accidental or unlawful destruction, loss, alteration, unauthorised disclosure of, or access to, personal data transmitted, stored or otherwise processed'.[13]

---

9 Article 4(1) of the GDPR.

10 The European Court of Justice says "the preamble to a Community act has no binding legal force and cannot be relied on either as a ground for derogating from the actual provisions of the act in question or for interpreting those provisions in a manner clearly contrary to their wording" (Case C-136/04 *Deutsches Milch-Kontor GmbH v Hauptzollamt Hamburg-Jonas* [2005] ECR I-10095, [32]). See also Tadas Klimas and Jurate Vaiciukaite, 'The Law of Recitals in European Community Legislation' (2008) 15 ILSA Journal of International & Comparative Law 61.

11 EDPB, 'Home' <https://edpb.europa.eu> accessed 26 June 2023.

12 See Commission, 'Article 29 working party archives 1997 – 2016' <https://ec.europa.eu/justice/article-29/documentation/index_en.htm> accessed 26 June 2023.

13 Article 4(12) of the GDPR. See in more detail: European Data Protection Board, Guidelines 9/2022 on personal data breach notification under GDPR, Version 2.0 Adopted 28 March 2023, 7-9 <https://edpb.europa.eu/our-work-tools/our-documents/guidelines/guidelines-92022-personal-data-breach-notification-under_en> accessed 26 June 2023.

Data breaches come in many forms. A controller can send one letter with personal data to the wrong addressee. If such a letter contains sensitive information, the breach can have serious effects for the data subject. Data breaches can also be larger. To illustrate: in 2013, it became known that names, birth dates, phone numbers and passwords of all Yahoo email users (3 billion, reportedly) were compromised.[14] The EDPB published guidelines with examples of types of data breaches, and suggestions to prevent such breaches and to mitigate harm if they occur.[15]

### 2.4 Who should notify?

The text of article 33 and 34 shows that the obligation to notify data breaches lies on the data controller. The GDPR defines a controller as the 'body which, alone or jointly with others, determines the purposes and means of the processing of personal data' (…).[16] A controller can be, for instance, a company, a municipality, or a university.

A controller can contract with a processor. A processor 'processes personal data on behalf of the controller.'[17] An example of a processor is a cloud storage provider that stores personal data for a controller. A processor is only allowed to do with personal data what the controller tells it to. A processor thus cannot use the personal data for its own purposes. The GDPR does not require processors to notify data breaches to the DPA or to the data subject. If a processor suffers a data breach, the processor must inform the relevant controller without undue delay about the breach.[18]

[14] Nicole Perlroth, 'All 3 Billion Yahoo Accounts Were Affected by 2013 Attack' (*New York Times*, 3 October 2017) <https://www.nytimes.com/2017/10/03/technology/yahoo-hack-3-billion-users.html> accessed 26 June 2023.

[15] EDPB, 'Guidelines 01/2021 on Examples regarding Personal Data Breach Notification', Adopted on 14 December 2021, Version 2.0.

[16] Article 4(7) of the GDPR.

[17] Article 4(8) of the GDPR.

The EU lawmaker could have required processors to notify data breaches to DPAs and data subjects too. It is debatable whether the lawmaker made the right choice on this point. Now, the processor must notify the controller; the controller must notify the data subject (in the case of high-risk breaches). It might be quicker if a processor directly notified the data subject. In some situations, data subjects can limit damage if they react quicker, for instance by blocking their credit cards or changing their password after a breach.

On the other hand, a data processor may not know who the data subjects are, and the data subjects may not recognise the data processor if the processor would notify. A notification from an unknown party may be confusing for a data subject. The chosen solution in the GDPR thus sacrifices notification time for a uniform reporting process and recognisable notification sender.

### 2.5 No-risk data breaches: no notification

If a controller suffers a data breach three situations can be distinguished: (i) non-risky data breaches do not have to be notified; (ii) if that non-risk exception does not apply, the DPA must be notified, and (iii) high-risk data breaches must be notified to the DPA and to the data subject. We discuss each situation below.

Situation (i) concerns no-risk data breaches. The controller does not have to notify anybody about such data breaches. Article 33 says that the controller must notify all data breaches to the DPA, 'unless the personal data breach is unlikely to result in a risk to the rights and freedoms of natural persons'.[19]

---

18 Article 33(2) of the GDPR.

19 Article 33(1) of the GDPR.

When is a data breach 'unlikely to result in a risk to the rights and freedoms of natural persons'? An example mentioned by the EDPB is the loss of a sufficiently encrypted device, while the decryption key remains in sole possession of the controller and the device was not the sole copy of the data.[20]

Because the device is sufficiently secure and there is no loss of availability, no notification is required. If, however, it later becomes clear that the encryption is vulnerable, the breach may have to be reported after all.[21] This example illustrates why it is important for controllers to document all data breaches. Apart from that, the GDPR requires controllers to document all data breaches.[22]

## 2.6 Normal-risk data breaches: notification to DPA

A second possibility is that a controller suffers a data breach, and the exception for non-risky data breaches does not apply. Hence, the controller must notify the DPA. In the words of article 33: 'In the case of a personal data breach, the controller shall without undue delay and, where feasible, not later than 72 hours after having become aware of it, notify the personal data breach to the supervisory.'[23] If the controller does not notify the DPA within 72 hours, the controller must explain the reasons for the delay.'[24]

The GDPR's 72-hour notification window is quite short compared to, for instance, the 30 to 60-day timeframe that many US statutes offer.[25] While it is

---

20 EDPB (n 13) 19. See about 'encryption safe harbours': Mark Burdon, 'The conceptual and operational compatibility of data breach notification and information privacy laws' (DPhil thesis, Queensland University of Technology 2011). <https://eprints.qut.edu.au/47512/1/Mark_Burdon_Thesis.pdf> accessed 26 June 2023.

21 EDPB (n 13) 19.

22 Article 33(5) of the GDPR. See also section 2.9 of this paper.

23 See also recitals 85, 87, and 88 of the GDPR.

24 Article 33(1) of the GDPR. The EDPB says that 'without undue delay' means 'as soon as possible': EDPB (n 13) 20.

25 Most states data breach notification laws have a 45-day notification timeframe, while Florida has only 30-days. HIPAA, which covers health data, has a 60-day notification timeframe laudable to want to notify data subjects as soon as possible after a data breach, controllers need to time to investigate the depth and extent of the breach, and assess possible harms. This is particularly the case when the breach involves a complicated hack or an insider. Based on the US data (in particular the published notification letters), companies on average took the maximum amount of time that the law allowed them before notifying data subjects.[26]

A controller could notify the DPA in two stages.[27] That enables controllers to report to the DPA that a breach had occurred within 72 hours, while providing more details once the controller has gathered them.

### 2.7 High-risk data breaches: notification to data subject

High-risk data breaches must be notified not only to the DPA, but also to the relevant data subjects. In the words of article 34: 'When the personal data breach is likely to result in a high risk to the rights and freedoms of natural persons, the controller shall communicate the personal data breach to the data subject without undue delay.'[28] As the GDPR's preamble notes, a data breach can have far-reaching effects for people, and can lead, for instance, to identity fraud, financial loss, damage to reputation, and other privacy harms.[29]

---

regarding breach incidents. See: Fabio Bisogni, 'Information availability and data breaches: Data breach notification laws and their effects' (DPhil thesis, Delft University of Technology 2020), 163; Mahmood Sher-Jan, 'From incident to discovery to breach notification: Average time frames' (iapp, 26 September 2017) <https://iapp.org/news/a/from-incident-to-discovery-to-breach-notification-average-timeframes/> accessed 26 June 2023.

26 Importantly, the discovery period (the time it takes before a breach is detected) can be much longer, sometimes taking up to several months before a hack or insider theft is detected.

27 Article 33(4) of the GDPR.

28 Article 34(1) of the GDPR. See also recitals 75, 76, 86, 87, and 88 of the GDPR.

29 Recital 85 of the GDPR. See on the harms for data subjects: Dennis Gibson and Clive Harfield, 'Amplifying victim vulnerability: Unanticipated harm and consequence in data breach notification policy' (2022) IRV <https://journals.sagepub.com/doi/full/10.1177/02697580221107683> accessed 26 June 2023.

When is a data breach likely to lead to 'high risk'? Generally, the risk for people will be higher when a data breach concerns 'special categories of data' (sometimes called sensitive data), such as data revealing ethnic origin, political opinions, religious beliefs, data concerning health, sex life, or sexual orientation.[30] The EDPB mentions other factors to consider when assessing risk, such as the 'nature, sensitivity, and volume of personal data' and the 'severity of the consequences for individuals'.[31]

If a controller did not notify the relevant data subjects, the DPA can decide that the controller should notify them.[32] This possibility for a second opinion by the DPA provides more protection to the data subjects involved.

### 2.8 What should be included in the notification?

The GDPR describes what the notification to the DPA should include at a minimum. For instance, the notification should describe the nature of the data breach, including, where possible, the number of data subjects concerned, and the likely consequences of the breach.[33]

The GDPR requires 'plain language' in the notification to data subjects.[34] The notification must suggest measures that the data subject can take to mitigate negative effects of the breach, such as changing passwords.[35]

### 2.9 Obligation to maintain records of data breaches

Article 33 requires controllers to 'document any personal data breaches,

---

[30] Article 9(1) of the GDPR.
[31] EDPB (n 13) 23-26.
[32] Article 34(4) of the GDPR
[33] Article 33(3) of the GDPR. The EDPB provides more details: EDPB (n 13) 20-23.
[34] Article 34(2) of the GDPR.
[35] Article 33(3)(d) of the GDPR. See also EDPB (n 13) 20.

comprising the facts relating to the personal data breach, its effects and the remedial action taken.'[36] The GDPR adds that this documentation enables the supervisory authority to verify compliance with article 33.[37] The obligation to maintain records of data breaches is related to the GDPR's accountability principle, which says that controllers are responsible for, and must be able to demonstrate compliance with, the GDPR.[38] DPAs can fine controllers 10 million Euro, or up to 2 % of the total worldwide annual turnover, for failing to document data breaches.[39]

## 3 The rationales for notification: how does the GDPR fare?

A number of or goals of, or rationales for, data breach notification obligations can be identified. We identified these rationales by analysing the GDPR, EU policy documents, and literature.[40] The rationales are as follows. A data breach notification

(1) enables people to protect themselves after they receive a notification;
(2) enables people to choose or switch to competing services;
(3) can incentivise organisations to improve security;
(4) enables regulators to perform their duties, enforce the law and provide help;

---

36 Article 33(5) of the GDPR.
37 Article 33(5) of the GDPR.
38 Article 5(2) of the GDPR.
39 Article 83(4)(a) of the GDPR.
40 Nieuwesteeg and Faure discuss three goals ('social benefits'): Bernold Nieuwesteeg and Michael Faure, 'An analysis of the effectiveness of the EU data breach notification obligation' (2018) 34(6) CLSR 1232; Verstraete and Zarsky discuss four goals ('normative justifications'): Mark Verstraete and Tal Zarsky, 'Optimizing Breach Notification' (2021) 2021(3) U. Ill. L. Rev. 803; Barcelo and Traung discuss five 'purposes' of the notification obligation in the 2009 ePrivacy Directive: Barcelo and Traung (n 6).

(5) improves transparency and accountability;

(6) helps to compile statistics about data breaches.

Below, we explore for each rationale whether the GDPR's breach notification obligation is likely to be useful.

### 3.1 People can protect themselves

One rationale for a breach notification obligation is that people can protect themselves after a notification.[41] As the GDPR's preamble puts it, the controller should notify high-risk breaches to the data subject 'to allow him or her to take the necessary precautions.'[42]

Such an obligation could indeed be helpful in some cases. For instance, people can change their passwords if they are informed that a controller leaked their passwords, or block their credit card after a data breach involving the leak of credit cards.

However, several caveats are in order. First, some personal data are difficult to change. One's medical record can contain sensitive and high-risk data, but there is not much that people can do, even if they are aware that their medical data are leaked.[43] Second, imposing a notification obligation, while better than nothing, places the responsibility for protecting the data subject with the data subject him or herself, rather than the controller.

Third, there are reasons for scepticism regarding the effectiveness of enabling data subjects to protect themselves. For example, people often lack the technical knowledge to protect themselves properly against identity fraud and other risks.[44] Moreover, even if people had the technical knowledge to protect themselves, they may not protect themselves in practice. For instance, people can be subject to overconfidence. An example of overconfidence is speeding while driving: people tend to overestimate the time savings while underestimating the increased risk.[45] People may also underestimate the risks of a data breach.[46] Perhaps people's overconfidence could be mitigated by clearly framing the risks in a data breach notice.

---

41 Verstraete and Zarsky (n 40) 813-817, 817-823, 830-834; Gibson and Harfield (n 29) 6-7.

42 Recital 86 of the GDPR.

43 See also Solove and Hertzog (n 1) 45.

Numerous studies have pointed to the positive role that intermediary companies can play in protecting end users.[47] In this context, intermediaries are organisations that provide the infrastructure and platforms and enable communications and transactions between third parties and services.[48] Examples of intermediaries include internet service providers, search engines, credit rating agencies, payment system providers, etc.

Perhaps intermediaries could also play a useful role with handling the effects of data breaches. If banks or credit rating agencies are notified about large scale breaches, they can take proactive steps to help data subjects, for instance by being alert for fraudulent transactions. And there are websites where people can

44 Ross Anderson, 'Why information security is hard – an economic perspective' (Seventeenth Annual Computer Security Applications Conference, New Orleans, LA, USA, 10-14 December 2001) 358 <https://ieeexplore.ieee.org/document/991552> accessed 26 June 2023.

45 Eyal Peer, 'The time-saving bias, speed choices and driving behavior' (2011) 14(6) Transport. Res. Part F: Traffic Psychol. Behav. 543.

46 Alessandro Acquisti et. al, 'Nudges for privacy and security: Understanding and assisting users' choices online' (2017) 50(3) ACM Computing Surveys <https://dl.acm.org/doi/10.1145/3054926> accessed 26 June 2023.

47 Hadi Asghari, (2016) 'Cybersecurity via Intermediaries: Analyzing Security Measurements to Understand Intermediary Incentives and Inform Public Policy' (DPhil thesis, Delft University of Technology 2016).

48 Karine Perset, 'The Economic and Social Role of Internet Intermediaries' (2010) OECD Digital Economy Papers No. 171 <https://www.oecd-ilibrary.org/docserver/5kmh79zzs8vb-en.pdf?expires=1685805131&id=id&accname=ocid56023174a&checksum=49A9E40763FA22145EB3B85FDEA18B5B> accessed 26 June 2023.

subscribe, so the website informs the subscribers when their email address is included in a data breach about which the website has learned.[49]

On the other hand, when intermediaries themselves become too much involved in the processing of personal data, they themselves become an attractive target for criminals. This is particularly a concern for intermediaries that act as an identity provider, where users have a single account at this identity provider with which they subsequently sign in to their accounts at various other services. A data breach at this identity provider leaking login credentials then immediately compromises all linked accounts at these other services. [50] In sum, in some cases, the GDPR could be useful to help people protect themselves, but we shouldn't have overly optimistic exceptions.

### 3.2 People can choose, or switch to, competing services

A publicly known data breach could encourage customers to switch to a competitor. Kwon and Johnson found empirical evidence (in the US) that when a hospital suffered a data breach and alternatives were available, the hospital lost a significant number of patients in the long term.[51]

However, in many situations, this switching argument does not hold up. If your employer or university suffers a data breach, you cannot easily switch to another job or university.[52] And with many online services, switching is also difficult. For many online services, the usefulness depends on the number of other users.[53] For example, a messaging app or a social network app is only useful if your contacts also use the same app. If all your friends use Instagram, you will not have much fun if you alone move to a privacy-friendly social network service. As Shapiro and Varian note, '[w]hen the costs of switching from one brand of technology to another are substantial, users face *lock-in*.'[54] And nowadays, many online services are not interoperable.[55] For instance, you cannot send a direct message from Twitter to Facebook.

[49] One site where such dumps are searchable is: <https://haveibeenpwned.com/Passwords> accessed 26 June 2023.

[50] Gergely Alpár et. al, 'The Identity Crisis - Security, Privacy and Usability Issues in Identity Management' (2013) 9(1) Journal of Information System Security 23.

[51] Juhee Kwon and M. E. Johnson, 'The Market Effect of Healthcare Security: Do Patients Care about Data Breaches?' (14th Annual Workshop on the Economics of Information Security, WEIS 2015, Delft, The Netherlands, 22-23 June 2015).

[52] See also Solove and Hertzog (n 1) 45.

Even for services that are interoperable, such as email, switching to another provider takes time and effort. Suppose that an online email provider suffers a data breach. Even if the provider enables people to export their emails to import them into another service, many people might find such a procedure too difficult or time-consuming. Apart from that, people may overestimate the difficulty of switching.

Moreover, for most people it is difficult to assess whether another controller offers better security, even if data breaches are reported from time to time. There is information asymmetry regarding the security practices of controllers. Since the 1970s, economists have researched markets where customers cannot assess the quality of services or products: markets with information asymmetry. Nobel laureate Akerlof identified the lemons problem.[56] In a thought experiment, Akerlof used the used car market as an example of a market with information asymmetry. Suppose that bad used cars ('lemons') and good used cars are offered for sale. Sellers know whether they are offering a bad or a good car for sale, but a prospective buyer cannot identify hidden defects. Let's assume that all sellers say that they sell a good car, even if it's actually a lemon. A rational prospective buyer will choose the cheapest car, as the buyer does not know which car is actually of better quality than the others. As a result, someone who wants to sell a good used car will not be offered enough money. Sellers of good cars will therefore not offer their cars for sale. As a result, the average quality of second-hand cars on the market decreases. Buyers will therefore offer even lower prices. Therefore, fewer and fewer people will offer their cars for sale. The quality of cars offered for sale is decreasing further. Hence, in a market characterised by asymmetric information, sellers do not compete on quality. This can lead to low quality products and services.

---

53 Hal Varian, *Intermediate Microeconomics - A Modern Approach, New York* (W. W. Norton and Company 1999); Anderson (n 44) 358-365.

54 Carl Shapiro and Hal Varian, *Information Rules: A Strategic Guide to the Network Economy* (Harvard Business School Press 1999), 104.

55 Christopher T. Marsden and Ian Brown, *Regulating code: Good governance and better regulation in the information age* (MIT Press 2013).

56 George A. Akerlof, 'The Market for "Lemons": Quality Uncertainty and the Market Mechanism' (1970) 84(3) Quarterly Journal of Economics 488.

We admit that the lemons problem is a simplified hypothetical. In real life, many sellers honestly say if they are selling a low-quality car. Moreover, in real life, buyers are often protected by the law against hidden defects. And in many countries, car sellers are required to show a safety report or a vehicle test report. Nevertheless, the lemons problem illustrates how information symmetry in a market can lead to low quality products.

With many people not having sufficient time and technical knowledge to assess the security policies of controllers, there seems to be a lemons problem regarding privacy and data security.[57] There may thus be little competition in offering good data security, because customers have a hard time assessing the security practices of companies. Moreover, it can be rational (in the economic sense of the word) for consumers not to educate themselves. When the cost of learning is higher than the potential benefit derived from the decision, a rational consumer will not learn.[58]

---

[57] See also: Tony Vila et. al, 'Why We Can't Be Bothered to Read Privacy Policies Models of Privacy Economics as a Lemons Market' (ICEC '03: Proceedings of the 5th international conference on Electronic commerce, Pittsburgh Pennsylvania USA, 30 September 2003 – 3 October 2003) 403; Anderson (n 44).

The above outlines the problems encountered by a hypothetical rational consumer. However, people do not always decide rationally (in the economic sense of the word). Research shows that people structurally act differently than rational choice theory predicts.[59] People have limited mental resources to evaluate all possible options and consequences of their actions.[60] Because of their bounded rationality, people often rely on rules of thumb, or heuristics. A heuristic can be defined, in the words of Kahneman, as 'a simple procedure that helps find adequate, though often imperfect, answers to difficult questions.'[61] Most of the time, such mental shortcuts work fine. But heuristics can also lead to decisions that people later regret. Systematic deviations from rational choice theory, or common mistakes, are called biases.

One relevant bias for this paper is the 'status quo bias', or inertia.[62] This bias refers to the power of the default. Few people tweak the settings of their apps or their social network site accounts.[63] In line with the status quo bias, many people are not inclined to switch service providers quickly.

Empirical studies suggest that whether customers switch or not after a data breach are influenced by the recovery actions and communications of the organisation following the breach. Masuch et al. suggest, based on a survey study, that a company can keep consumers more loyal by expressing remorse and offering compensation.[64] Liang and Telang investigated the reactions of Home Depot's approximately 54 million customers whose banking details were exposed in 2014 in the US. The researchers suggest that steps taken by Home Depot were enough to convince most users not to switch, even when there were competing stores in close proximity. However, these steps cost the firm $300 million.[65]

---

58 Alessandro Acquisti and Jens Grossklags, 'What Can Behavioral Economics Teach Us About Privacy?' in Alessandro Acquisti et al. (eds), *Digital Privacy: Theory, Technologies and Practices* (Taylor and Francis Group 2007).

59 Cass R. Sunstein, 'Introduction' in Cass R. Sunstein (ed), *Behavioral law and economics* (CUP 2000), 1.

60 Herbert A. Simon, *Models of man: social and rational; Mathematical Essays on Rational Human Behavior in Society Setting* (Wiley 1957).

61 Daniel Kahneman, *Thinking, Fast and Slow* (Allen Lane 2011), 98.

62 William Samuelson and Richard Zeckhauser, 'Status Quo Bias in Decision Making' (1988) 1 Journal of Risk and Uncertainty 7.

63 See e.g., Alessandro Acquisti et. al, 'Privacy and Behavioral Economics' in Bart P. Knijnenburg et. al, *Modern Socio-Technical Perspectives on Privacy* (Springer 2022).

To conclude, in theory, people can switch to a competing service after a notification of a data breach. However, we do not think that this effect of a data breach notification obligation should be exaggerated. In many cases, people find it difficult or burdensome to switch. Even if the costs of switching are minimal, people might not switch in practice. Regardless of our criticism, notification obligations can be useful for those consumers who would like to switch to another company as a result of a data breach.

### 3.3 Incentivising organisations to improve security

A third rationale for data breach notification obligations is that such obligations can push organisations towards better security.[66] A notification obligation could stimulate controllers to focus on data security, as reported data breaches cause negative publicity. Indeed, a fear of fines and reputation damage may cause organisations to invest in security. Research from the US suggests that organisations paid more attention to data security after a notification obligation law came into force.[67]

---

64 Kristin Masuch et. al, 'Do I Get What I Expect? An Experimental Investigation of Different Data Breach Recovery Actions' (Twenty-Eighth European Conference on Information Systems (ECIS2020), An Online AIS Conference, 15-17 June 2020) <https://aisel.aisnet.org/ecis2020_rp/37/> accessed 26 June 2023.

65 Yangfan Liang and Rahul Telang, 'Customer Response to Adverse Security Events: An Empirical Study' (15 January 2020) <https://papers.ssrn.com/sol3/papers.cfm?abstract_id=3523788> accessed 26 June 2023.

66 See Verstraete and Zarsky (n 40) 813-817; Gibson and Harfield (n 29) 7-8.

In Europe too, it appears that controllers started to take data security more seriously after the GDPR was adopted in 2016. Especially the GDPR may be able to influence the behaviour of controllers, as the GDPR's contains a possibility of high fines.[68] We think that the GDPR, and data breach notification obligations in general, indeed incentivises controllers to improve their data security.[69] However, it would be difficult to distinguish the effects of the GDPR in general, and the effects of its data breach notification requirements.

Murcian-Goroff provides some evidence on the long-term effects of breach notification obligations in the US.[70] Using archival data, he reconstructs when approximately 214,000 companies updated their web server software, and applied necessary security patches, in the years before and after California passed a breach notification law. By comparing companies within and outside of the jurisdiction of California, he finds that firms headquartered in California updated their server software between one to four weeks earlier than other firms, thus reducing the possibility of a breach.

---

67 Samuelson Law, Technology and Public Policy Clinic, 'Security Breach Notification Laws: Views from Chief Security Officers' (2007) Technical report, University of California, Berkeley <www.law.berkeley.edu/files/cso_study.pdf> accessed 26 June 2023.

68 Article 83 of the GDPR.

69 Cisco, 'Maximizing the value of your data privacy investments: Data Privacy Benchmark Study' (2019) Cisco Cybersecurity series <https://www.cisco.com/c/dam/en_us/about/doing_business/trust-center/docs/dpbs-2019.pdf> accessed 26 June 2023.

70 Raviv Murciani-Goroff, 'Do Data Breach Do Data Breach Disclosure Laws Increase Firms' Investment in Securing Their Digital Infrastructure?' (18th Annual Workshop on the Economics of Information Security (WEIS 2019), Boston, Massachusetts, USA, 3-4 June 2019) <https://weis2019.econinfosec.org/wp-content/uploads/sites/6/2019/05/WEIS_2019_paper_33.pdf> accessed 26 June 2023.

The impacts of breach notification obligations are, as one might expect, different based on the economic sector. In a study of 202 severe US security breaches between 2005 and 2019, Malliouris and Simpson found that severe security breaches are generally associated with increased 'market risk' for the company – meaning it becomes more expensive for these firms to obtain money from investors.[71] However, industrial firms (who have less direct contact with consumers) are less prone to an increase in market risk after a breach.

An obligation to notify individuals of a data breach can incentivise controllers to improve their security in another way. People can sue a controller if they suffered damage from a data breach.[72] Controllers could invest in security because they fear such court cases, in which a court might order the controller to pay damages to victims. As Nieuwesteeg and Faure note, 'Liability results in behaviour that incentivizes organizations to internalize some of the externalities in cyber security.' [73]

However, only the future can tell how important this liability argument will be in practice. Compared to the US, it is rare in Europe that victims of a data breach sue a controller.[74] Going to court is burdensome and expensive. And it can be difficult to prove monetary damage after a data breach. Many judges in Europe tend not to award large damages after GDPR violations.[75] There are examples of collective action procedures in which groups of people sue a controller for GDPR violations.[76] It is too early to say how successful such cases will be, and to what extent the cases will incentivise controllers to improve compliance and security.

---

71 Dennis D. Malliouris and Andrew Simpson, 'Underlying and consequential costs of cyber security breaches: Changes in systematic risk' (16th Annual Workshop on the Economics of Information Security (WEIS 2017), University of California San Diego, USA, 26-27 June 2017) <https://weis2017.econinfosec.org/wp-content/uploads/sites/9/2020/06/weis20-final14.pdf> accessed 26 June 2023.

72 Nieuwesteeg and Faure (n 40) section 3.2.

73 Ibid.

74 See also Angela Daly, 'The introduction of data breach notification legislation in Australia: A comparative view' (2018) 34(3) CLSR 477, 488.

75 See for instance Timotheus Franciscus Walree, 'Schadevergoeding bij de onrechtmatige verwerking van persoonsgegevens' ['Compensation for unlawful processing of personal data'] (DPhil thesis, Radboud University Nijmegen 2021), English summary on pp. 221-224 <https://hdl.handle.net/2066/231874> accessed 26 June 2023.

### 3.4 The obligation enables regulators to perform their functions

A fourth rationale for a data breach notification obligation is that the obligation enables DPAs to do their job. The requirement to report a personal data breach to the DPA has a number of functions.

First, the DPA can provide a second opinion if a controller decided not to notify the data subject. If the DPA considers the personal data breach to be of high risk, then the DPA may require the controller to communicate the breach to the data subject, regardless of the opinion of the controller.[77]

Second, the notifications give DPAs information about security risks in the society. For instance, a DPA may find that many data breaches occur in a certain sector, or that one type of data breach (say: stolen unencrypted laptops) regularly occur. In reaction, the DPA could, for example, warn or educate controllers, or focus investigations on a certain sector.

Verstraete and Zarsky note that a controller that does not take security seriously could be lucky and not be breached. Another controller can suffer a data breach because of bad luck, while it did invest in security.[78] Data Protection Authorities should consider this factor when considering fining data controllers.

---

[76] See Federica Casarosa, 'Transnational collective actions for cross-border data protection violations' (2020) 9(3) Internet Policy Review <https://policyreview.info/articles/analysis/transnational-collective-actions-cross-border-data-protection-violations> accessed 26 June 2023.

[77] Article 34(1) of the GDPR.

[78] Verstraete and Zarsky (n 40).

### 3.5 Improving transparency and accountability

A fifth rationale for data breach notification obligations is that openness about data breaches is in line with data protection law's transparency principle.

A data breach notification obligation can help to improve transparency in several ways. First, a data breach notification obligation can contribute to transparency towards the data subject.[79] The first of the GDPR's overarching principles emphasises transparency: personal data must be 'processed lawfully, fairly and in a transparent manner in relation to the data subject'.[80] Case law confirms the importance of transparency.[81] Mitigating information asymmetry could be seen as one of the main goals of the GDPR.[82] In the light of data protection law's transparency principle, it makes sense to inform the data subject about data breaches.

There is a second way in which a data breach notification obligation can contribute to transparency. Controllers themselves may learn from documenting their breaches, as the GDPR requires. They may also learn from having to notify their breaches. Both requirements may lead to better understanding of security risks.

Third, as noted in the previous section, if organisations notify DPAs about data breaches, the DPAs learn more about how common breaches are, in which sectors there are the most data breaches, etc. Moreover, DPAs could compile statistics about data breaches; we discuss that in the next section.[83]

---

79 Nieuwesteeg & Faure speak of a 'right to know': Nieuwesteeg and Faure (n 40) section 3.2.

80 Article 5(1)(a) of the GDPR.

81 See for instance: Case C-201/14 *Smaranda Bara and Others v Casa Naţională de Asigurări de Sănătate and Others* [2015] ECLI:EU:C:2015:638, [33]; ECtHR, Grand Chamber, *Barbulescu v Romania* App no 61496/08 (ECtHR, 5 September 2017), [133].

82 See Paul De Hert and Serge Gutwirth, 'Privacy, Data Protection and Law Enforcement. Opacity of the Individual and Transparency of Power' in Erik Claes et. al (eds), *Privacy and the Criminal Law* (Intersentia 2006).

There is, however, the question of whether the GDPR's transparency requirements go far enough, especially when compared with data breach notification laws in other jurisdictions. Bisogni investigated how the differences among data breach notification laws across the United States impacted the notifications letter sent to data subjects.[84] His analysis is made possible by the fact that several US states publicize the contents of breach notification letters to affected individuals. The GDPR does not contain such a requirement. Bisogni recommends that in addition to the information already required to be given to data subjects by the GDPR, notifications should give the breach detection date, and highlight the possibility to receive customer support regarding the breach if applicable. He further proposes notifying credit agencies, as in his analysis expanding the visibility of notifications reduces information asymmetries. Notification of credit agencies, however, might be more important in the US than in Europe, due to the fact that credit cards are less commonly used in Europe.

In sum, the GDPR's obligation to notify data breaches contributes to transparency. However, as discussed in the next section, the GDPR could have contributed more to transparency.

### 3.6 Generating statistics

A data breach notification obligation enables authorities to compile and publish statistics about data breaches. The EDPB already publishes some data: it sometimes reports on the number of data breaches that have been notified to the

[83] As Solove and Hartzog note about the US: 'Before 2005, there were certainly many data breaches, but companies weren't required to report them, so there isn't much recorded history.', Solove and Hertzog (n 1) 18, 44.

[84] Bisogni (n 25) 151-176.

DPAs.[85]

But more transparency by the EDPB and DPAs would be useful.[86] We recommend that European DPAs and the Board make more statistics and information available for researchers and others. To illustrate: the Attorney Generals of several US states publish details about data breaches, including the breached firm or organisation's name and even the notification letters sent to data subjects.[87] The cybersecurity research community has gained a lot of knowledge from analysis of incident data. In particular the US data breach reports have inspired numerous research papers that have studied region- and sector-specific policies, high profile incidents, the impact of investments, and the various consequences of incidents. Such research can contribute to better security, and to more informed policymaking.

As the Dutch Cyber Security Council (an independent advisory body) notes:

> Information on data breaches is an important source of insight into the actual effects of security measures (or a lack thereof). Better research into the effects of such breaches enables organisations to take better decisions as to the security measures in which they should invest. This in turn may increase demand for cyber insurance, allowing insurers to mandate a higher basic level of security measures as part of their policy terms.[88]

---

85 EDPB, 'Contribution of the EDPB to the evaluation of the GDPR under Article 97' (18 February 2020) <https://edpb.europa.eu/our-work-tools/our-documents/other/contribution-edpb-evaluation-gdpr-under-article-97> accessed 26 June 2023.

86 Bisogni (n 25) 3, 165.

87 As of 2019, 22 states require notifications to the Attorney General, but only 10 states publish details of the events and the notification letters. These ten states include California, Indiana, Maine, Maryland, Montana, New Hampshire, Oregon, Vermont, Washington, and Wisconsin. See Bisogni (n 25).

88 Dutch Cyber Security Council, CSR Advisory document 'Making data breach notifications available for research purposes' (2020) CSR Advisory document 2020, No. 1, 2

Making available data does not necessarily mean a full open data policy. For instance, the identity of the breached controller can remain hidden in many situations. And a tiered system could be developed, where researchers and others can get different types of access, depending on the type of research they want to do. There are many shades between not publishing data at all, and making data completely open.[89]

Building on the data published by the EDPB, we calculate the number of breach notifications per hundred thousand inhabitants and per hundred thousand firms, for all EU member states (except Greece). We provide the results in table 1.

---

<https://www.cybersecuritycouncil.nl/advisory-documents/documents/advisory-documents/2020/02/11/csr-advisory-document-‘making-data-breach-notifications-available-for-research-purposes’---csr-advisory-document-2020-no.-1> accessed 26 June 2023.

89 Frederik Zuiderveen Borgesius et. al, ‘Open data, privacy, and Fair Information Principles: towards a balancing framework’ (2015) 30(3) Berkeley Technology Law Journal 2073.

Table 1—breach notifications by country, per hundred thousand inhabitants/enterprises

| **Country** | **Breach Notifications**[90] | **Active Enterprises**[91] | **Population**[92] | **Breaches per 100K enterprises** | **Breaches per 100k population** |
|---|---|---|---|---|---|
| Austria | 1,536 | 202,946 | 8,858,775 | 756.9 | 17.3 |
| Belgium | 1,220 | 193,947 | 11,455,519 | 629.0 | 10.6 |
| Bulgaria | 98 | 182,854 | 7,000,039 | 53.6 | 1.4 |
| Croatia | 115 | 105,889 | 4,076,246 | 108.6 | 2.8 |
| Cyprus | 92 | 33,852 | 875,899 | 271.8 | 10.5 |
| Czechia | 665 | 223,532 | 10,649,800 | 297.5 | 6.2 |
| Denmark | 9,418 | 123,323 | 5,806,081 | 7,636.9 | 162.2 |
| Estonia | 171 | 57,756 | 1,324,820 | 296.1 | 12.9 |
| Finland | 5,762 | 135,281 | 5,517,919 | 4,259.3 | 104.4 |
| France | 3,273 | 1,149,541 | 67,012,883 | 284.7 | 4.9 |
| Germany | 45,561 | 1,505,451 | 83,019,213 | 3,026.4 | 54.9 |
| Greece | not reported | 284,564 | 10,724,599 | - | - |
| Hungary | 744 | 382,522 | 9,772,756 | 194.5 | 7.6 |
| Iceland | 340 | 15,978 | 356,991 | 2,127.9 | 95.2 |
| Ireland | 9,500 | 114,713 | 4,904,240 | 8,281.5 | 193.7 |
| Italy | 1,951 | 1,385,602 | 60,359,546 | 140.8 | 3.2 |
| Latvia | 153 | 72,778 | 1,919,968 | 210.2 | 8.0 |
| Lithuania | 268 | 75,737 | 2,794,184 | 353.9 | 9.6 |
| Luxembourg | 498 | 20,447 | 613,894 | 2,435.6 | 81.1 |

90 EDPB (n 85) 35.

91 Eurostat, 'Structural Business statistics - Business demography - Employer business demography by size class (bd_9fh_sz_cl_r2). Data from 2017 for firms with ≥ 1 employee' <https://ec.europa.eu/eurostat/web/structural-business-statistics/data/database> accessed 26 June 2023.

92 Eurostat, 'Demography and Migration - Population - Population on 1 January (tps00001), Data from 2019' <https://ec.europa.eu/eurostat/web/population-demography-migration-projections/data/main-tables> accessed 26 June 2023.

| | | | | | |
|---|---|---|---|---|---|
| Malta | 218 | 13,350 | 493,559 | 1,633.0 | 44.2 |
| Netherlands | 37,413 | 242,673 | 17,282,163 | 15,417.0 | 216.5 |
| Norway | 2,587 | 130,441 | 5,328,212 | 1,983.3 | 48.6 |
| Poland | 7,957 | 770,375 | 37,972,812 | 1,032.9 | 21.0 |
| Portugal | 400 | 316,876 | 10,276,617 | 126.2 | 3.9 |
| Romania | 521 | 406,325 | 19,414,458 | 128.2 | 2.7 |
| Slovakia | 158 | 130,269 | 5,450,421 | 121.3 | 2.9 |
| Slovenia | 193 | 66,887 | 2,080,908 | 288.5 | 9.3 |
| Spain | 1,434 | 1,332,290 | 46,937,060 | 107.6 | 3.1 |
| Sweden | 6,794 | 292,527 | 10,230,185 | 2,322.5 | 66.4 |
| U. Kingdom | 21,000 | 2,317,245 | 66,647,112 | 906.2 | 31.5 |

The EDPB report highlights a large variance in the number of data breaches across Europe. To make the numbers comparable, we need to account for the size of each country and economy, for instance, by calculating the ratio of breaches per 100,000 firms in each country. The result ranges from under 200 (Italy, Spain, Romania) to over 7,000 (Denmark, Ireland, Netherlands) breach notifications per 100,000 firms in the specified time period. Such a large difference is somewhat puzzling. A higher ratio does not necessarily reflect a worse security situation; it can also indicate better breach detection capabilities, and a more cautious business culture that prefers to over-report.

In sum, a data breach notification obligation can help authorities to compile information and statistics. However, in the EU, more could be done to make such information available for stakeholders.

## 4 Discussion and suggestions for further research

We must mention some limitations of this paper. It is notoriously difficult to empirically assess the effect of regulation. And at the moment, there are little empirical data that prove whether the GDPR contributes to the six goals of the data breach notification obligations. As noted, the DPAs make only limited data available.

There are ample opportunities for more research into data breach notification obligations. Several empirical questions warrant further research. For example, we reported on differences in reported data breaches across Europe.[93] How could these differences be explained? And can the effect of data breach notification obligation laws be measured? Which effects should ideally be measured? In the US, researchers found that while baseline disclosure requirements have small effects, requiring firms to notify state regulators reduces identity fraud report rates by roughly 10%.[94] Bisogni proposes that to have similar studies in the EU, a centralized public repository is necessary to track identity fraud and data breach information across different EU countries.[95] It would also be interesting if more research is published on how people react to data breach notifications, and to what extent they take measures to protect themselves.[96]

Regarding legal research, several topics could be explored. Solove and Hartzog distinguish three categories of data security laws, roughly summarised: (i) breach notification requirements; (ii) laws with security requirements; (iii) laws that assign liability and enable private litigation.[97] The GDPR, and EU law more generally, contain rules in category (i) and (ii). But perhaps policymakers could do more to enable private litigation regarding data breaches. Other open questions include: Could policymakers or regulators do more to minimise the harm of data breaches, once they have occurred?[98] Should the EDPB give more detailed guidance on the content of notifications to data subjects?[99]

---

93 See section 3.6.

94 Aniket Kesari, 'Do Data Breach Notification Laws Work?' (5 August 2022, last revised 30 August 2022) <https://papers.ssrn.com/sol3/papers.cfm?abstract_id=4164674> accessed 26 June 2023; See also Sasha Romanosky et. al, 'Do data breach disclosure laws reduce identity theft?' 30(2) JPAM 256. See also Bisogni (n 25) chapter 6.

95 Bisogni (n 25) 151-176.

96 Gibson and Harfield (n 29) 19.

97 Solove and Hertzog (n 1) 37.

New types of voice assistants and smart speakers also trigger legal questions. Examples include Google Home, Amazon's Alexa, and Apple's HomePod. Such devices often have a single registered account, while they may have multiple users, such as family members. Voice recordings from such devices can contain all sorts of information, including personal data, such as addresses, telephone numbers, and similar data. If a provider of such devices suffers a data breach, the provider may only have the contact information of the account holder. It is not immediately apparent how the provider could notify the other users. The device may even have recorded information about somebody who visited the house with the 'smart' speaker. The GDPR says that a controller does not have to notify all data subjects if that would involve a disproportionate effort.[100] In such a case, a controller could use a public notification, such as an advertisement in a widely-read medium.[101] Whether that solution is sufficient requires more research.

## 5 Conclusion

We analysed the GDPR's rules on data breaches. We identified six main rationales for data breach notification obligations. The rationales are as follows: (i) after a notification, people can protect themselves (ii) and can switch to

98 See Solove and Hertzog (n 1) 195; Bisogni (n 25) 165.

99 Bisogni (n 25) 162. The EDPB did give some guidance on how to mitigate harm from data breaches. EDPB (n 15).

100 Article 34(3)(c) of the GDPR.

101 See also EDPB (n 13) 21.

competing services; (iii) a data breach notification obligation can incentivise organisations to improve security; (iv) enables regulators to perform their duties; (v) contributes to transparency and accountability; (vi) and can help to generate statistics for researchers, policymakers, and others.

We summarised the main findings from literature (law, security economics, empirical research) to assess whether the GDPR is likely to contribute to these goals. We argued that the GDPR can, in principle, contribute to these goals. However, we also warn that we should not have exaggerated expectations of the possibilities for people to protect themselves after a data breach. For instance, if embarrassing pictures of somebody are leaked online, the victim has limited possibilities to stop the spread of those pictures. Likewise, we should not have high expectations of people switching to other service providers after receiving a data breach notification. Switching costs time, and people may find it difficult. Moreover, it is difficult for people to assess whether competitors offer better security.

Lastly, we call upon Data Protection Authorities to publish more information about data breaches that have been reported to them. Such information enables research and better policymaking regarding data security.